\documentclass[twocolumn,showpacs,preprintnumbers,amsmath,amssymb]{revtex4}

\usepackage{graphicx}
\usepackage{dcolumn}
\usepackage{bm}

\begin{document}

\title{Diffusion entropy analysis on the stride interval fluctuation of human gait}
\author{Shi-Min Cai$^{1}$}
\author{Pei-Ling Zhou$^{1}$}
\author{Hui-Jie Yang$^{2}$}
\author{Tao Zhou$^{1,2}$}
\email{zhutou@ustc.edu}
\author{Bing-Hong Wang$^{2}$}
\author{Fang-Cui Zhao$^{3}$}

\affiliation{
$^{1}$Department of Electronic Science and Technology, University of Science and Technology of China, Hefei Anhui, 230026, PR China\\
$^{2}$Department of Modern Physics, University of Science and Technology of China, Hefei Anhui, 230026, PR China\\
$^{3}$College of Life Science and Bioengineering, Beijing
University of Technology, Beijing, 100022, PR China }

\date{\today}

\begin{abstract}
In this paper, the diffusion entropy technique is applied to
investigate the scaling behavior of stride interval fluctuations
of human gait. The scaling behavior of the stride interval of
human walking at normal, slow and fast rate are similar; with the
scale-invariance exponents in the interval $[0.663,0.955]$, of
which the mean value is $0.821\pm0.011$. Dynamical analysis of
these stride interval fluctuations reveals a self-similar pattern:
Fluctuation at one time scale are statistically similar to those
at multiple other time scales, at least over hundreds of steps,
while the healthy subjects walk at their normal rate. The
long-range correlations are observed during the spontaneous
walking after the removal of the trend in the time series with
Fourier filter. These findings uncover that the fractal dynamics
of stride interval of human gait are normally intrinsic to the
locomotor systems.
\end{abstract}


\maketitle

\section{Introduction}
Recently, it has been recognized that in many natural sequences
the elements are not positioned randomly, but exhibit long-range
correlations and fractal dynamics. Prominent examples include
noncoding DNA sequences \cite{Peng1994}, human heartbeat
\cite{Ivanov1999,Bunde2000}, human brain electroencephalogram
\cite{Zheng2003}, teenagers' births in Texas \cite{Scafetta2004},
and financial time series \cite{Yamasaki2005}. The common feature
of all these diverse systems with long-range correlations is that
scaling behavior decay by a power law, where a characteristic
scale is absent, and the scaling behavior is usually rather robust
and universal.

The scaling behavior in complex systems is not only interesting in
physical sense, but also provides an intrinsic description of the
system, and highlights the system dynamical mechanism. Several
useful variance-based methods, such as the probability moment
method \cite{Paladin1987} and the fluctuation approach as well as
the de-trended fluctuation approach \cite{Peng1994}, are proposed
to detect the scale-invariance properties. However, these
variance-based methods have two basic shortcomings. One is that
the scale-invariance property can be detected but the value of the
exponent cannot be obtained accurately. The other is that for some
processes, like the L\'{e}vy flight, the variance tends to
infinite and these methods are unavailable at all
\cite{Scafetta2001,Grigolini2001}. Although the infinite can not
be reached due to the finite records of empirical data, clearly we
will obtain misleading information about the dynamics under these
conditions.

The dynamics of human gait is relative to locomotor system's
synthesizing inputs from the motor cortex, the basal ganglia and
the cerebellum, as well as feedback from the vestibular, visual
and proprioceptive sources. Therefore, to investigate the scaling
behavior of the stride interval fluctuations of human gait is very
interesting with both the physical and physiological communities.
By using some recently proposed approaches for nonlinear data,
some scientists studied the statistical characters of the stride
interval fluctuations of human gait
\cite{Hausdorff1995,Hausdorff1996,West1998,West1999,Griffin2000,Goldberger2002,Costa2003,Perc2005}.
Hausdorff \emph{et al} \cite{Hausdorff1995,Hausdorff1996}
demonstrated that the stride interval time series exhibits
long-range correlations, suggesting that the human gait displays a
self-similar activity. Subsequent studies, by West \emph{et al}
and Goldberger \emph{et al}
\cite{West1998,West1999,Griffin2000,Goldberger2002}, supported the
above conclusion, and several dynamics models are established to
mimic the human gait \cite{Ashkenazy2002,West2003}. Furthermore,
Perc \cite{Perc2005} found that the human gait process possesses
some typical properties like a deterministic chaotic system. In
this paper, we apply the diffusion entropy (DE) technique to
accurately detect and obtain the scaling property of stride
interval fluctuations of human gait on a Fourier analysis.

\begin{figure}
\scalebox{0.8}[0.8]{\includegraphics{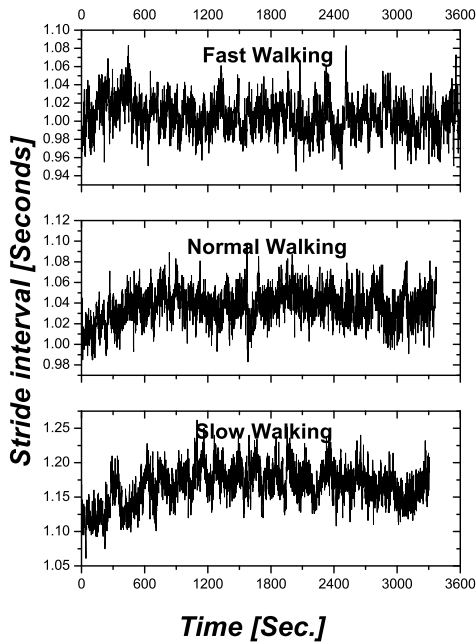}}
\caption{\label{fig:epsart} A typical example about the stride
interval fluctuation of human gait.}
\end{figure}

\section{diffusion entropy technique based on Fourier analysis}
To overcome the mentioned shortcomings in the variance-based
methods, Grigolini \emph{et al.} \cite{Scafetta2001,Grigolini2001}
designed the approach named diffusion entropy analysis (DEA). To
keep our description as self-contained as possible, we briefly
review the DEA method.

Consider a complex system containing a large amount of particles.
The scale-invariance property in the diffusion process of this
system can be described mathematically with the probability
distribution function as
\begin{equation}
P(x,t)=\frac{1}{t^\delta}F(\frac{x}{t^\delta}),
\end{equation}
where $x$ is the displacements of the particles in the complex
system and $\delta$ the scale$-$invariance exponent. The
theoretical foundation of this property is the Central Limit
Theorem and the Generalized Central Limit Theorem
\cite{Ma1985,Gnedenko1954}. For $\delta=0.5$, the diffusion
process is the standard diffusion and $F(y)$ is the Gaussian
function. And $\delta\neq0.5$ indicates the deviation of the
dynamical process from the normal stochastic one. For example, the
case with $\delta>0.5$ corresponds to the diffusion process of
L\'{e}vy walk.

For a time series, the delay-register vectors, denoted with
$\{y_{k},y_{k+1},\cdot\cdot\cdot,y_{k+m-1}|k=1,2,3,\cdot\cdot\cdot,N-m+1\}$,
can be regarded as the trajectories of $N-m+1$ particles during
the period of $0$ to $m$. In this way we can map a time series to
a diffusion process, called overlapping diffusion process here. An
alternative solution is to separate the considered time series
into many non-overlapping segments and regard these segments as
the trajectories.

To make the data suitable for the illustration of scaling behavior
at all scales, we propose a method based on the Fourier transform
to detrend the nonstationary trend in whole temporal domain.
First, transform the time series to Fourier space and then, use
the invert Fourier transform to obtain the time series with cutoff
coefficient. We consider the curtailed time series as the trend
and subtract it from original series. Then, the final time series
is regarded as a steady series, whose overlapping diffusion
process reads
\begin{equation}
x_k(t)=\sum^{k+t}_{j=k}\zeta_j,k=1,2,\cdots,N-t+1.
\end{equation}
Consequently, the Shannon entropy can be defined as
\begin{equation}
S(t)=-\int^{+\infty}_{-\infty}P(x,t)\log_{10}[P(x,t)]dx.
\end{equation}
A simple algebraic leads to
\begin{equation}
S(t)=A+\delta\log_{10}(t),
\end{equation}
where
\begin{equation}
A=-\int^{+\infty}_{-\infty}F(y)\log_{10}[F(y)]dx,y=\frac{x}{t^\delta}.
\end{equation}
The DEA method has been used to analysis many time series in
different research fields, such as the solar induced atmosphere
temperature series \cite{Grigolini2002}, the intermittency time
series in fluid turbulence \cite{Bellazzini2003}, the spectra of
complex networks \cite{Yang2004}, the output spike trains of
neurons \cite{Yang2005}, the index of financial market
\cite{Cai2006}, and so on.

\section{Data Analysis}
Herein, we map a time series to a diffusive process and introduce
the DE technique to investigate the time series of stride interval
fluctuations of human gait, which is obtained from healthy
subjects who walked for 1 hour at their normal, slow and fast
paces \cite{Physionet}. The data contains the stride interval
fluctuations of ten young healthy men, given an arbitrary ID
$(si01, si02, si03, \cdots, si10)$. Participants have no history
of any neuromuscular, respiratory, or cardiovascular disorders and
are taking no medication. Mean age is 21.7 yr (range $18-29$ yr).
Height is $1.77\pm0.08 \texttt{(SD)} m$, and weight is
$71.8\pm10.7  \texttt{(SD)} kg$. All the subjects provided
informed written consent. Subjects walked continuously on level
ground around an obstacle free, long (either 225 or 400 meters),
approximately oval path and the stride interval is measured using
ultra-thin, force sensitive switches taped inside one shoe. A
typical example is shown in Fig. 1.

The stride interval significantly decreased and the velocity
significantly increased with each chosen walking rate, as
expected. The mean stride intervals of the three subjects are
$1.3\pm0.2s$, $1.1\pm0.1s$, and $1.0\pm0.1s$ during the slow,
normal, fast walking trials, respectively. And the mean velocities
are $1.0\pm0.2m/s$, $1.4\pm0.1m/s$, and $1.7\pm0.1m/s$ during the
slow, normal, and fast walking, respectively. The mean velocity
increases by an average of $77\%$ from slow to fast walking. A
wide range of walking rates are obtained, enable us to test for
the effects of walking rate on the scaling behavior indicating the
long-range correlations.

The locomotor control system maintains the stride interval at an
almost constant level throughout the 1 hour walking. Nevertheless,
the stride interval fluctuations is in a highly complex, seemingly
random fashion. In order to truly uncover the scaling behavior of
the stride interval fluctuation of human gait, we study all the 30
samples classifying into three subjects: Fast, normal, and slow.
The scaling behaviors of the three classes of subjects, obtained
by using DE analysis, are shown in Fig. 2A, 2B and 2C,
respectively. Each figure contains 10 samples, of which the time
scale is from 10 to 300. The results indicate that the stride
interval time series is not completely random (uncorrelated),
instead, it exhibits the scale-invariance property and long-range
correlation at all the three walking rates.

\begin{figure}
\scalebox{0.7}[0.6]{\includegraphics{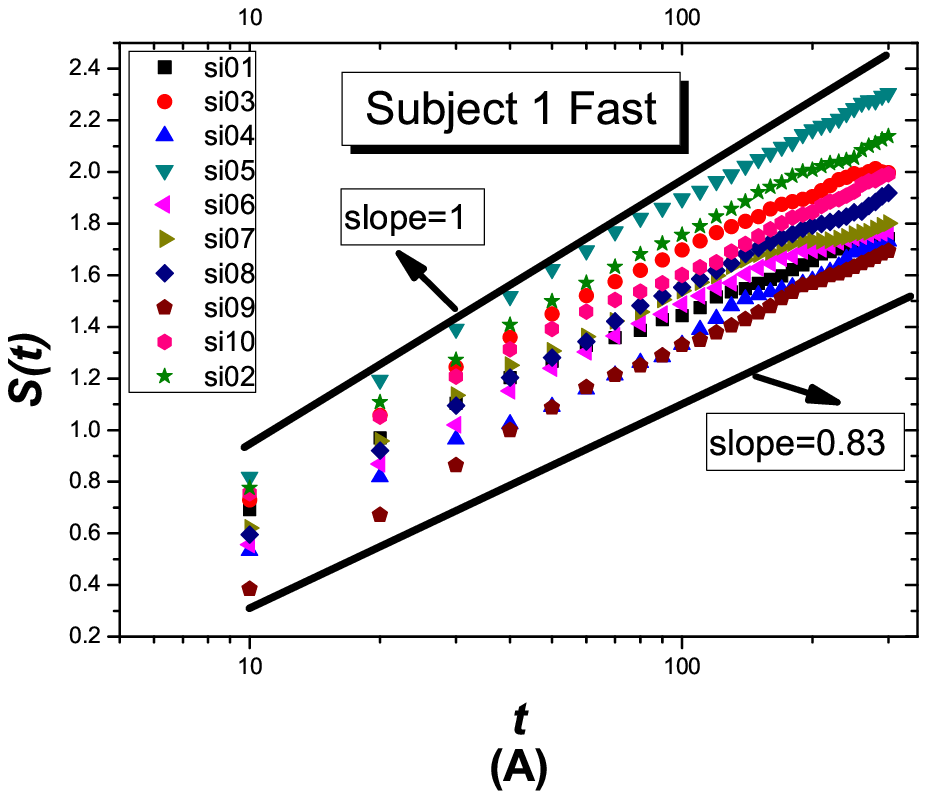}}
\scalebox{0.7}[0.6]{\includegraphics{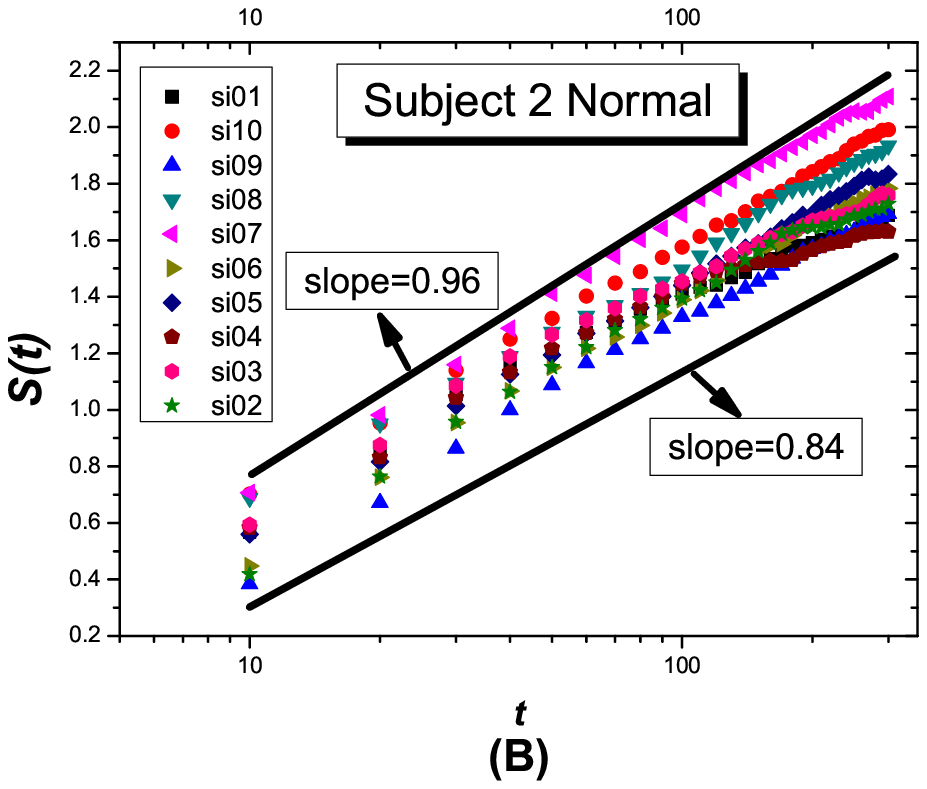}}
\scalebox{0.7}[0.6]{\includegraphics{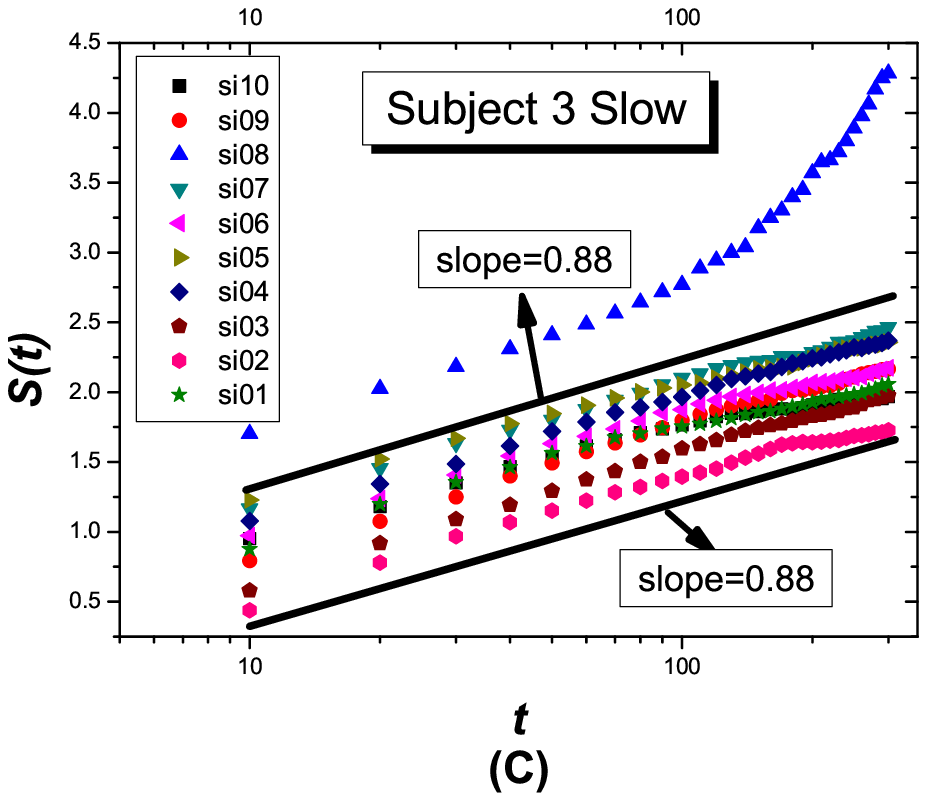}}
\caption{\label{fig:epsart} (Color online) The scaling behavior of
the stride interval fluctuations of human gait. (A, (B) and (C)
denote the the subject Fast, Normal and Slow respectively.}
\end{figure}

Furthermore, Fig. 3 illustrates the dependence of $\delta$ on
self-determined walking rate for all three subjects. For the 29
samples of 30 1-h trials, except one samples $si08$ of the slow
subject (see Fig.2C), the scale-invariant exponents $\delta$ is
around $0.821\pm0.011$ (range 0.663 to 0.955). Thus for all
subjects at all rates, the stride interval time series displayed
scaling behavior and long-range power law correlations.

\begin{figure}
\scalebox{0.8}[0.8]{\includegraphics{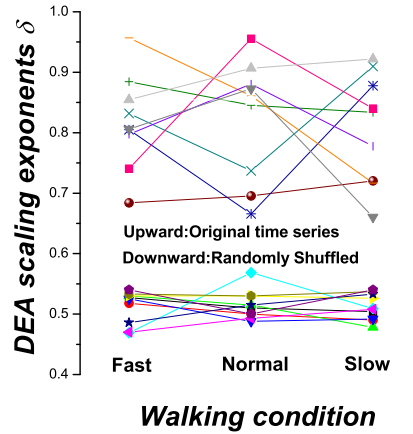}}
\caption{\label{fig:epsart} (Color online) Dependence of $\delta$
 on self-selected walking rate.}
\end{figure}

There are a strong similarity of $\delta$ on chosen walking rate:
$\delta$ is $0.829\pm0.011$, $0.827\pm0.011$, $0.809\pm0.013$
during the fast-, normal-, slow-walking trails, respectively (see
Fig. 3). In principle, the estimation of $\delta$ depends on the
fitting range $(t1,t2)$. It is preferable to have $t_2\gg t_1$.
However, due to the finite size of data, a larger $t_2$ may lead
to a bigger error in $\delta$ \cite{Peng1993}. In this paper, all
the results are obtained by linear fitting within the region
$(10,300)$.

\section{Conclusion}
In summary, by means of the DE method we investigate the scaling
behavior embedded in the time series of stride interval
fluctuation. Scale-invariance exponents of the three subjects are
almost the same, being in the interval of $[0.663, 0.955]$, of
which the mean value is $0.821\pm0.011$. Dynamical analysis of
these step-to-step fluctuations reveals a self-similar pattern:
Fluctuation at one time scale are statistically similar to those
at multiple other time scales, at least over hundreds of steps,
while the healthy subjects walk at their normal rate. The
long-rang correlation is observed during the spontaneous walking
by removal of the drift or trend in the time series. Thus the
above features uncover the fractal dynamics of spontaneous stride
interval are normally intrinsic to the locomotor system.

\begin{acknowledgements}
This work was partially supported by the National Natural Science
Foundation of China under Grant Nos. 70471033, 10472116, 10532060,
10547004, 70571074, and 70571075, the Special Research Founds for
Theoretical Physics Frontier Problems under Grant No. A0524701,
and Specialized Program under the Presidential Funds of the
Chinese Academy of Science.
\end{acknowledgements}


\begin{thebibliography}{Peng1994}
\bibitem{Peng1994} C. -K. Peng, S. V. Buldyrev, S. Havlin, M. Simons, H. E. Stanley, and A. L. Goldberger, Phys. Rev. E 49 (1994) 1685.
\bibitem{Ivanov1999} P. C. Ivanov, L. A. N. Amaral, A. L. Goldberger, S. Havlin, M. G. Rosenbulm, Z. R. Struzik and H. E. Stanley, Nature 399 (1999) 461.
\bibitem{Bunde2000} A. Bunde, S. Havlin, J. W. kantelhardt, T. Penzel, J. H. Peter and K. Voigt, Phys. Rev. Lett. 85 (2000) 3736.
\bibitem{Zheng2003}  C. P. Pan, P. Zheng, Y. Z. Wu, Y. Wang and X. W. Tang, Physica A 329 (2004) 130.
\bibitem{Scafetta2004} N. Scafetta and B. J. West, Chao, Solitons and Fractals 20 (2004) 179.
\bibitem{Yamasaki2005} K. Yamasaki, L. Muchnik, S. Havlin, A. Bunde and H. E. Stanley, Proc. Natl. Acad. Sci. U.S.A. 102 (2005) 9424.
\bibitem{Paladin1987} G. Paladin and A. Vulpiani, Phys. Rep. 156 (1987) 147.
\bibitem{Scafetta2001} N. Scafetta, P. Hamilton, and P. Grigolini, Fractals 9 (2001), 193.
\bibitem{Grigolini2001} P. Grigolini, L. Palatella, and G. Raffaelli, Fractals 9 (2001), 439.
\bibitem{Hausdorff1995} J. M. Hausdorff, C. -K. Peng, Z. Ladin, J. Y. Wei and A. L. Goldberger, J. Appl. Physiol. 78 (1995) 349.
\bibitem{Hausdorff1996} J. M. Hausdorff, P. L. Putrdon, C. -K. Peng, Z. Ladin, J. Y. Wei and A. L. Goldberger, J. Appl. Physiol. 80 (1996) 1448.
\bibitem{West1998} B.J. West and L. Griffin, Fractals 6(1998) 101.
\bibitem{West1999} B.J. West and L. Griffin, Chaos, Solitons and Fractals 10 (1999) 1519.
\bibitem{Griffin2000} L. Griffin, D. J. West and B. J. West, J. Biol. Phys. 26 (2000) 185.
\bibitem{Goldberger2002} A. L. Goldberger, L. A. N. Amaral, J. M. Hausdorff, P. C. Ivanov, C. K. Peng and H. E. Stanley, Proc. Natl. Acad. Sci. U.S.A. 99 (2002) 2466.
\bibitem{Costa2003} M. Costa, C. K. Peng, A. L. Goldberger and J. M. Hausdorff, Physica A 330 (2003) 53.
\bibitem{Perc2005} M. Perc, Eur. J. Phys. 26 (2005) 525.
\bibitem{Ashkenazy2002} Y. Ashkenazya, J. M. Hausdorff, P. C. Ivanova and H. E. Stanley, Physica A 316 (2002) 662.
\bibitem{West2003} B. J. West and N. Scafetta, Phys. Rev. E 67 (2003) 051917.
\bibitem{Ma1985} S. K. Ma, Statistic Mechanics, World Scientific, Singapore, 1985.
\bibitem{Gnedenko1954} B. V. Gnedenko and A. N. Klomogorove, Limit Distributions for Sum of Independence Random Variables, Addison Wesley, Reading, 1954.
\bibitem{Grigolini2002} P. Grigolini, D. Leddon, and N. Scafetta, Phys. Rev. E 65 (2002) 046203.
\bibitem{Bellazzini2003} J. Bellazzini, G. Menconi, M. Ignaccolo, G. Buresti, and P. Grigolini, Phys. Rev. E 68 (2002) 026126.
\bibitem{Yang2004} H. -J. Yang, F. -C. Zhao, L. Qi, and B. -L. Hu, Phys. Rev. E 69 (2004) 066104.
\bibitem{Yang2005} H. -J. Yang, F. -C. Zhao, W. Zhang, and Z. -N. Li, Physica A 347 (2005) 704.
\bibitem{Cai2006} S. -M. Cai, P. -L. Zhou, H. -J. Yang, C. -X. Yang, B. -H. Wang, and T. Zhou, Physica A 367 (2006) 337.
\bibitem{Physionet} See http://physionet.org/physiobank/database/umwdb/.
\bibitem{Peng1993} C. -K. Peng, S. V. Buldyrev, A. L. Goldberger, S. Havlin, M. Simons and H. E. Stanley, Phys. Rev. E 47 (1993) 3730.
\end{thebibliography}
\end{document}